# Adaptive Model for Computer-Assisted Assessment in Programming Skills

P. Molins-Ruano, C. González-Sacristán, F. Díez, P. Rodriguez and G. M. Sacha
Departamento de Ingeniería Informática. Escuela Politécnica Superior. Universidad Autónoma de Madrid
Campus de Cantoblanco, 28049 Madrid, Spain
{pablo.molins, carlos.gonzalez }@estudiante.uam.es, {fernando.diez, pilar.rodriguez, sacha.gomez}@uam.es

In this work, we show a methodology aimed to improve the quality of the assessment process for subjects related to basic programming. The method takes into account the relevance of the items and the students' answers to follow different paths to improve the accuracy of the assessment process. We have developed numerical simulations and experiments with real students that demonstrate the advantages of this model when compared with traditional evaluation tools. This method improves the objectiveness and takes into account the relevance of the subject contents. We also demonstrate that the architecture of the algorithm is fully compatible with traditional multiple choice test formalisms. Our results can be directly used in computer-assisted tests for different subjects and disciplines, as well as used by the students as a self-evaluation tool with the objective of correcting their deficiencies in the learning process.

## 1. INTRODUCTION

Being essential for the students' performance of their potential professional activities, it is more than adequate to consider the use of computers in any dimension of their learning process. One of the topics that can be strongly improved by computer-assisted tools is the assessment process. In this context, Computer-based testing (CBT) can offer distinct advantages. Within CBT, Computer Adaptive Testing (CAT) has been used for very different purposes such as instantaneous scoring [1], language proficiency improvement [2], learning styles identification [3], measurement of chess playing proficiency [4], Maths ability [5], improvement of the efficiency of personality [6], or heath status assessment [7]. Different versions of CAT have been used in learning environments in order to personalize the learning process [8] [9] [10] [11]. In that framework, CAT allows seeing the students as individuals, taking their own characteristics into account. Typically, CAT systems are able to adapt the items presented to the student depending on their former answers, often including some kind of personalized feedback [12].

One of the most important difficulties in CAT systems is the need of pre-calibrating the items before they are used in real assessments. For this reason, the test should be performed by a large sample of the population in order to get a good test calibration [5]. The complexity of pre-calibration has led to a huge interest in the development of students' models which could help in the pre-calibration task [13] [14] [15] [16].

Here, we propose a computer-assisted method for subjects related to basic programming that takes into account the relevance of the items in the path followed by the assessment process. The goal of this method is to increase the objectivity, robustness and relevance of the contents. To be sure that the assessment process is adequate, the model should not lose any relevant information about the students that can be extracted from traditional evaluation tools such as open answer tests.

In section 2, we describe our model and show numerical simulations for both multiple choice and completion type items. As we will demonstrate, the model shows a more robust and efficient behavior when completion type items are used. The completion test item is a free response type of item in which the student must supply the missing information. Completion type is a kind of test commonly used in language learning environments, where a missing word in a sentence could be the answer. In programming environments, we make the students provide the output of an algorithm performance given certain input data. In section 3, we show results from real experiments where we show that the use of completion type items is appropriate in subjects related to programming skills. As we will demonstrate by numerical simulations and experiments with real students, our model will be a very useful tool with great advantages for their use in the assessment of programming skills.

## 2. ASSESSMENT MODEL

To take into account the relevance of the items we must design a test model able to give better assessments to the students with a higher understanding of the most basic (i.e. relevant) topics of the subject. Although the model will allow for any number of levels, here we have only considered three in order to simplify the simulations. Moreover, for practical reasons, considering a low value of levels would require a simpler classification process. The first level includes the most important and basic concepts of the subject. As levels go up, the items focus on specific concepts which are only relevant after getting a robust knowledge of the questions posed in previous levels.

We propose a computer assisted model that stresses the importance of the answers in lower levels. The model is motivated by the need of making a clear difference between basic and advanced knowledge. The model follows different paths depending on the students' answers, which make essential the use of appropriate software packages to correctly configure the next step of any item. For example, some engines initially designed for gamification could be directly applied by taking advantage of the conversational rules; where different students follow different paths depending on their answers in a conversation with a non-playable character [17].

In this model, all students should answer the same number of questions (although from different levels depending on their answers). The model is defined by two parameters: the number of levels for the items $N_l$ and the number of questions that must be answered by the student $N_v$. Once we have defined those numbers, the items are placed as shown in figure 1, where the item level increases from left to right, being constant top-down. The test begins at the question placed on the top. A wrong answer (light arrow) will move one step down to the next question in the same column. A correct answer (dark arrow) will move one step down and also one step to the right. Since the item level does not change vertically, the only way to start answering questions belonging to a different level is to answer a certain number of questions correctly (i.e. move rightwards). For example, if an exam has been conceived to include $N_l=3$ $N_v=36$, every student should answer at least 12 questions to reach a new level. The scores are shown at the bottom, being the highest ones at the right side. The number of different grades in this model is $N_v+1$.

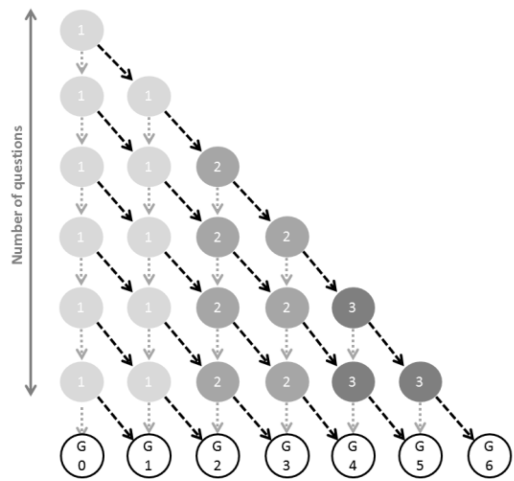

Fig 1. Model 1 graphical description. Black and gray arrows represent the directions followed by correct and wrong answers respectively. Numbers inside the nodes represents the item levels. White dots correspond to the end of the evaluation process, where the grades are obtained.

The scores placed at the bottom of the figure must be defined in different ways depending on the kind of questions of the test. For simplicity, let us define the scores S between 0 (lowest) and 1 (highest). For completion tests, it is easy to see that fair scores can be obtained by dividing the grades shown in the figure between 0 and 1. In the case of multiple choice tests, we must be careful about the fact that students have a non-zero probability of choosing the correct answer randomly. This fact is well-known in other multiple choice tests, and formula scoring [18] is a widely used procedure. In the case of students that answer all the items, the scores are obtained by the following expression:

$S=(N_a*C -1)/(N_a-1)$                                       (1)

Where C is the number of items answered correctly divided by $N_v$, S is the final score (between 0 and 1) and $N_a$ is the number of possible answers per item. It is easy to demonstrate that the scores obtained by equation 1 can be easily fit in our model by assuming a linear distribution and a wise selection of the lowest and highest values in the final scores. This choice, however, must be taken carefully since equation 1 has been developed assuming that students have the option of leaving items with no response. This choice is not possible in our model since the path of the test depends on the students' answers.

To test our model, we have performed several numerical simulations with four kinds of student. The first one (good) is a student who takes into account questions from all the item levels. The second student (bad) is a student who does not achieve a good level in any kind of questions. The third student (direct) is the one who gives more importance to the basic knowledge (i.e. the first levels). In opposition, the fourth student (inverse) focuses on the most advanced questions without taking care of the basic ones. We have simulated our students with the probability of giving correct answers at any level. We have also defined the four students facing two different test formats: completion and multiple choice tests with three possible answers. In table 1 we include the probabilities for the eight possibilities that rise by combining the 4 students with the 2 kinds of tests.

A key condition here is that our model should give good and bad assessments for the good and bad students respectively. Additionally, direct and inverse students must be also distinguished by giving better assessments for the direct ones. It is worth noting that direct and inverse students have the same probabilities in different order. This is an important fact since direct and inverse students would have the same scores in tests which do not take into account the item level.

Simulations have been performed for $10^5$ students of each class. We have fixed the number of items to 36. Additionally, we have also considered the possibility of answering 3 items per node. In this scenario, we must also specify the minimum number of correct answers inside each node that makes the test follow the correct answer path. We have considered that the number of correct answers can be either 3/3 or 2/3. To keep the number of total questions constant, we have reduced the vertical levels to 12 when 3 questions are asked.

| Student | Test type | Level | | |
|---|---|---|---|---|
| | | 1 | 2 | 3 |
| Good | Completion | 90% | 80% | 70% |
| | Multiple Choice | 95% | 95% | 95% |
| Bad | Completion | 20% | 20% | 20% |
| | Multiple Choice | 33.33% | 33.33% | 33.33% |
| Direct | Completion | 90% | 50% | 10% |
| | Multiple Choice | 95% | 70% | 45% |
| Inverse | Completion | 10% | 50% | 90% |
| | Multiple Choice | 45% | 70% | 95% |

Table 1. Correct Answer probabilities for the four kinds of students, two test types and three levels for the items.

In figure 2 we show the grade distributions for the completion type tests. As we can see, good and direct students get good grades. Moreover, as we expected, bad and inverse students get bad grades. The direct student scores are not as good as the ones from the good student. However, the inverse student grades are comparable to the ones from the bad student. In this case, we do not find any advantage of having 3 questions per node (figures 2b and 2c). However, the number of different grades is drastically reduced by a factor 3, which makes these configurations much worse.

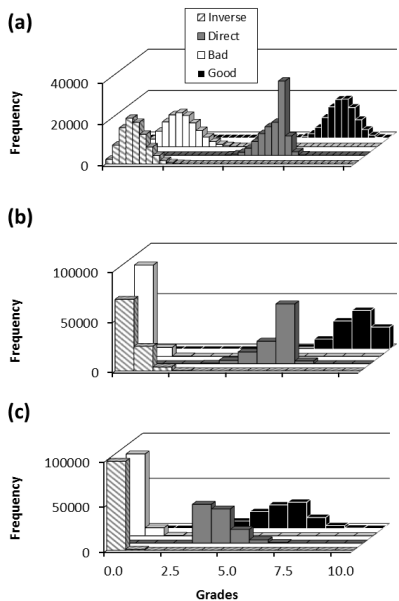

Fig 2. Grade distributions for model 1 and completion tests. (a) 36 levels with 1 question per node. (b) 12 levels with 2/3 correct answers per node. (c) 12 levels with 3/3 correct answers per node.

We find that the worst configuration is the one depicted in figure 2c, where 3/3 correct answers are needed per node. The problem here is that the good student distribution becomes tail weighted.

In figure 3 we show distributions for a multiple choice test with 3 possible answers. As we can see in that figure, the 36 levels configuration (3a) is much worse since the distributions for the bad and inverse student have higher values. In this case, we can see the advantages of including 3 items per node (figures 3b and 3c). When 3/3 correct answers are required, the distributions for the bad and inverse students go to the lower limit. At the same time, grades from direct students are clearly higher than those from bad and inverse, being still the ones from the good student at the top of the distribution.

From this model, for both completion type and multiple choice tests, we have found a configuration that matches our requirements. In the case of completion tests, the best configuration is the one that uses a single question per node. For multiple choice tests, the best option is 3 questions per node and consider a correct answer only when 3/3 have been correctly answered.

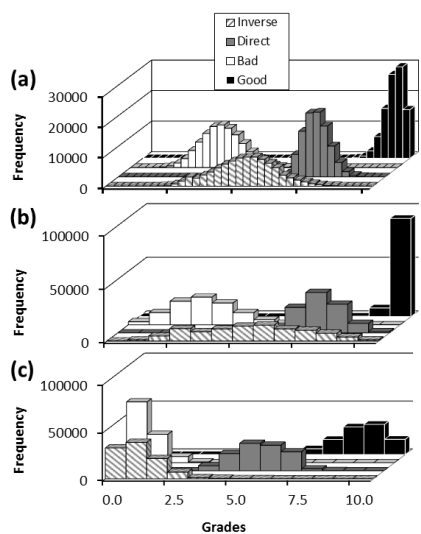

Fig 3. Grade distributions for model 1 and multiple choice tests. (a) 36 levels with 1 question per node. (b) 12 levels with 2/3 correct answers per node. (c) 12 levels with 3/3 correct answers per node.

In conclusion, we have found that completion tests are the most adequate to be included in our model using a single item per node. Now, we must be sure that these kinds of tests are adequate for the assessment of programming skills. This is the topic analyzed in the next section.

A free evaluation software that includes the model shown in this section is available online. [19]

### 3. COMPLETION TYPE TESTS IN PROGRAMMING SKILLS EVALUATION

This study has been made with students from the subject Computer Methods in the academic year 2012/2013. The choice of that subject is appropriate because the use of a subject that deals with basic knowledge simplifies the development of items for the tests, which is one of the most challenging tasks for the completion type tests developed here.

#### 3.1 Description of the subject "Computer Methods"

The subject Computer Methods is a compulsory course in the 1st year of Chemical Engineering Bachelor Degree at the Science Faculty in Universidad Autónoma de Madrid. About 100 students took this course in the 2012/2013 academic year, which was taught through theoretical and practical lessons. Each student attended 3 hours per week of theory and 2 hours in the computer room. This subject corresponds to 6 ECTS.

The contents of the subject are mainly related to fundamentals of Computer Science and Computer Programming in Matlab language. The methodology is based in master lectures followed by practices in computer sessions. Lessons include subjects related to problem analysis, pseudo code generation, exemplifications, choice of action and decision making, mostly related to Matlab programming.

In the initial assessments, the students' work was evaluated by means of grades based on two open answer final exams and some practical works. Final grades took in consideration both the numerical results and the quality of the code. The marking process took a long time, due to the requirement of evaluating the code in detail. Moreover, the evaluation of the open answer tests became difficult since there were many students who used different approximations to solve the programming exercises. The methodology shown in the following section corrects these problems in a very efficient way.

### 3.2 Completion type tests development

To include the completion type test in the assessment, we changed the model of the tests both for the theory and practical evaluations. In both cases, we used an algorithm able to generate questions that give a huge number of different possible answers by changing a few parameters. With this program, we created a number of different tests that equaled the number of students. The items were, however, virtually the same since the changes introduced by the algorithm did not change the difficulty or the process that students must follow to answer correctly. For example, a question that was used in the study was written as follows:

> "The following code calculates the definite integral of the function f(x)=cos(x) between -1 and 1, in 0.01 increments.
>
> ```
> ini=-1;
> fin=1;
> inc=0.01;
> inte=0;
> for x=ini:inc:fin
>     inte=inte+cos(x+inc/2);
> end
> inte=inte*inc;
> disp(inte)
> ```
>
> This algorithm must be modified so as to use the function $f(x)=x^2$ between 10 and 11, in 0.5 increments.
>
> Answer:______"

As we can see, this very simple question only asks the student to make small changes in the code. The student's task is to identify the sections of the algorithm that must be changed and make the correct modifications. The changes that our generator introduces here are only related to the limits 10 and 11 (another student would receive a test with the limits 5 and 6, for example). This difference does not change the difficulty of the item since the student's task is just to understand the code and change the correct numbers. However, the final answer is different, so students cannot copy this number from a different test. Questions arising from this kind of test can be much more difficult if we ask the students to make a bigger number of changes or even to make the whole code by themselves.

To study the effect of using completion tests in the assessments, we decided to design tests with the format shown before both for theory and practice. However, in the theory test we asked the students to give the final result and also describe the code used (open answer). In the practical test we only asked for the final number (pure completion test). The objective of this strategy was to check if the students' assessments were similar with completion and open answer tests.

We have done the study only with students who answered both kinds of tests. After removing the students who did not complete both tasks, we had 30 students from the Chemical Engineering grade along the 2012/2013 academic course. The students' grades are shown in figure 4. Results are ordered as a function of the grades obtained in the completion test (i.e. practical test). A good correlation is obtained between both grades. We did a statistical study of the 30 cases and found a Pearson correlation coefficient between both scores to be 0.66. The results are t=4.6487, p=0.0001. In the case of Spearman coefficient, we found r=0.623 and p=0.000233.

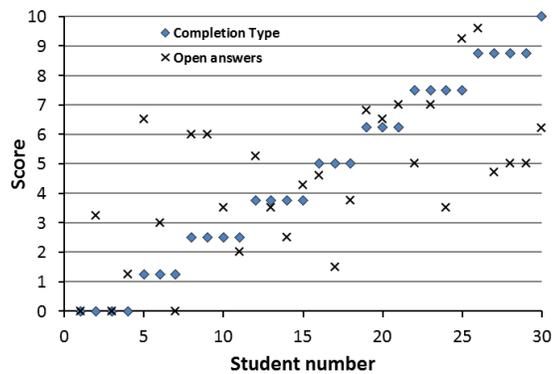

Fig 4. Students' scores from the tests with completion type and open answer formats.

## 4. CONCLUSION AND FUTURE WORK

We have presented a computer assisted model for the evaluation of programming skills that gives objective, complete and secure assessments. The main characteristic of this model is that it takes into account the relevance of the items, changing the path as a function of the students' answers. We have made numerical simulations that demonstrate that this model can be used for both completion type and multiple choice tests. However it gives a better performance for completion tests. This fact implies that completion tests are a better choice in terms of stability of the models. To be sure that completion tests are an adequate format to be used with our method, we have performed a study with real students that demonstrate that this kind of tests gives similar results than open answer tests where the quality of the code is judged.

We have found that the main advantages of our model are the following:

1) The use of different tests for any student avoids the possibility of cheating by copying the answers. This is a very interesting characteristic of our model because it gives an immense number of different tests without changing their difficulty, which is one of the main issues that must be carefully taken into account when different test models are being used.

2) The evaluation process becomes 100% objective since only the final result of the exercises is taken in consideration. It is also simplified since the evaluators only have to compare the numbers given by the students with the correct answer. This process can be also automatized and done by a computer.

3) We have found a very good correlation between open answer and completion type tests. This fact indicates that our completion test gives similar assessment than open answer tests, using a much simpler and objective evaluation process.

In conclusion, this study reveals that completion tests give objective, secure and complete assessments, without losing any relevant information when compared with open answer tests. This fact makes completion test items a good choice to be included in computer-assisted assessment models. Even, some self-evaluation tools can be developed to be used by the students. With these tools, students will be able to check whether they are focusing on the most relevant topics or not, and correct their deficiencies if needed.

## 5. ACKNOWLEDGMENTS

Authors acknowledge Sarah Dobber and Pablo Castells for their support and interesting discussions. This work has been partially founded by projects ASIES (TIN2010-17344), e-Madrid (S2009/TIC-1650) and (UAM/EPS-L2.6.13). GMS acknowledges support from the Spanish "Ramón y Cajal Program".